\documentclass[11pt,a4paper]{article}

\usepackage{jheppub,bm}

\title{Neutrino interaction with matter in a noninertial frame}

\author[a,b,c,d]{Maxim Dvornikov}

\affiliation[a]{Institute of Physics, University of S\~{a}o Paulo, \\
CP 66318, CEP 05315-970 S\~{a}o Paulo, SP, Brazil}
\affiliation[b]{Pushkov Institute of Terrestrial Magnetism, Ionosphere
and Radiowave Propagation (IZMIRAN), \\
142190 Troitsk, Moscow, Russia}
\affiliation[c]{Nonlinear Physics Centre, Research School of Physics and Engineering,
Australian National University, \\
2601 Canberra, ACT, Australia}
\affiliation[d]{Physics Department, Tomsk State University, \\ 634050 Tomsk, Russia}
\emailAdd{maxim.dvornikov@usp.br}

\abstract{
We study the system of massive and mixed neutrinos interacting with
background matter moving with an acceleration. We start with the derivation
of the Dirac equation for a single neutrino in the noninertial frame
where matter is at rest. A particular case of matter rotating with
a constant angular velocity is considered. The Dirac equation is solved
and the neutrino energy levels are found for ultrarelativistic particles
propagating in rotating matter. Then we generalize our results to
include several neutrino generations and consider mixing between them.
Using the relativistic quantum mechanics approach we derive the effective
Schr\"{o}dinger equation for the description of neutrino flavor oscillations
in rotating matter. We obtain the resonance condition for neutrino
oscillations and examine how it can be affected by the matter rotation.
We also compare our results with the findings of other authors who
studied analogous problem previously.
}


\arxivnumber{1408.2735}

\begin{document}

\maketitle

\section{Introduction}

Nowadays it is commonly believed that neutrinos are massive particles
and there is a nonzero mixing between different neutrino generations.
These neutrino properties result in transitions between neutrino flavors,
or neutrino oscillations~\cite{Bel14}. It is also known that various
external fields, like electroweak interaction of neutrinos with background
fermions~\cite{BleSmi13}, neutrino electromagnetic interaction~\cite{BroGiuStu12},
and neutrino interaction with a strong gravitational field~\cite{AhlBur96},
can also influence the process of neutrino oscillations.

As shown in ref.~\cite{Lam01}, noninertial effects in accelerated
and rotating frames can affect neutrino propagation and oscillations.
The consideration of the reference frame rotation is particularly
important for astrophysical neutrinos emitted by a rapidly rotating
compact star, e.g., a pulsar. For example, the possibility of the pulsar spin down
by the neutrino emission and interaction with rotating matter was
recently discussed in ref.~\cite{DvoDib10}. It should be noted that besides elementary particle physics, various processes in noninertial frames are actively studied in condensed matter physics. For instance, the enhancement of the spin current in a semiconductor moving with an acceleration was recently predicted in ref.~\cite{BasCho13}.

In the present work we shall study the neutrino interaction with background
matter in a rotating frame. We assume that neutrinos can interact
with background fermions by means of electroweak forces. We also take
that neutrino mass eigenstates are Dirac particles. In our treatment
we account for the noninertial effects since we find the exact solution
of a Dirac equation for a neutrino moving in the curved space-time with
a metric corresponding a rotating frame.

This work is organized in the following way. In section~\ref{sec:NUMATFLS},
we start with the brief description of the neutrino interaction with
background matter in Minkowski space-time. We consider matter moving
with a constant velocity and discuss both neutrino flavor and mass eigenstates.
Then, in section~\ref{sec:MASSNUNONIF}, we study background matter
moving with an acceleration. The Dirac equation, including noninertial
effects, for a neutrino interacting with such matter is written down
in a frame where matter is at rest. In section~\ref{sub:ROTATION},
we solve the Dirac equation and find the neutrino energy spectrum
for ultrarelativistic neutrinos moving in matter rotating with a constant
angular velocity. Then, in section~\ref{sec:OSC}, we apply our results
for the description of neutrino oscillations in rotating background
matter. The effective Schr\"{o}dinger equation governing neutrino oscillations
is derived and the new resonance condition is obtained. We consider
how the matter rotation can affect the resonance in neutrino oscillation
in a realistic astrophysical situation. Finally, in section~\ref{sec:SUMMARY},
we summarize our results and compare them with the previous findings
of other authors.

\section{Neutrino interaction with background matter\label{sec:NUMATFLS}}

In this section we describe the interaction of different neutrino
flavors with background matter in a flat space-time. We discuss a
general case of matter moving with a constant mean velocity and having
a mean polarization. Then we consider the matter interaction of neutrino
mass eigenstates, which are supposed to be Dirac particles.

The interaction of the neutrino flavor eigenstates $\nu_{\alpha}$,
$\alpha=e,\mu,\tau$, with background matter in the flat space-time
is described by the following effective Lagrangian~\cite{GiuKim07}:
\begin{equation}\label{eq:Lf}
  \mathcal{L}_{\mathrm{eff}} =
  -\sum_{\alpha}\bar{\nu}_{\alpha}\gamma_{\mu}^{\mathrm{L}}\nu_{\alpha}\cdot f_{\alpha}^{\mu},
\end{equation}
where $\gamma_{\mu}^{\mathrm{L}}=\gamma_{\mu}(1-\gamma^{5})/2$, $\gamma^{\mu}=(\gamma^{0},\bm{\gamma})$
are the Dirac matrices, and $\gamma^{5}=\mathrm{i}\gamma^{0}\gamma^{1}\gamma^{2}\gamma^{3}$.

The interaction Lagrangian in eq.~(\ref{eq:Lf}) is derived in the
mean field approximation using the effective external currents $f_{\alpha}^{\mu}$
depending on the characteristics of background matter as~\cite{DvoStu02}
\begin{equation}\label{eq:flavnueffp}
  f_{\alpha}^{\mu} =
  \sqrt{2}G_{\mathrm{F}}\sum_{f}
  \left(
    q_{\alpha,f}^{(1)}j_{f}^{\mu}+q_{\alpha,f}^{(2)}\lambda_{f}^{\mu}
  \right),
\end{equation}
where $G_{\mathrm{F}}$ is the Fermi constant and the sum is taken
over all background fermions $f$. Here
\begin{equation}\label{eq:jms}
  j_{f}^{\mu}=n_{f}u_{f}^{\mu},
\end{equation}
is the hydrodynamic current and
\begin{equation}\label{eq:lambdams}
  \lambda_{f}^{\mu} =
  n_{f}
  \left(
    (\bm{\zeta}_{f}\mathbf{u}_{f}),
    \bm{\zeta}_{f}+\frac{\mathbf{u}_{f}(\bm{\zeta}_{f}\mathbf{u}_{f})}{1+u_{f}^{0}}
  \right),
\end{equation}
is the four vector of the matter polarization. In eqs.~(\ref{eq:jms})
and~(\ref{eq:lambdams}), $n_{f}$ is the invariant number density
(the density in the rest frame of fermions), $\bm{\zeta}_{f}$ is
the invariant polarization (the polarization in the rest frame of
fermions), and $u_{f}^{\mu}=\left(u_{f}^{0},\mathbf{u}_{f}\right)$
is the four velocity. To derive eqs.~(\ref{eq:flavnueffp})-(\ref{eq:lambdams})
it is crucial that background fermions have constant velocity. Only
in this situation one can make a boost to the rest frame of the fermions
where $n_{f}$ and $\bm{\zeta}_{f}$ are defined.

The coefficients $q_{\alpha,f}^{(1,2)}$ in eq.~(\ref{eq:flavnueffp})
can be found in the explicit form if we discuss the system of active
neutrinos moving in matter composed of electrons $e$, protons $p$,
and neutrons $n$. In this situation, one gets for $\nu_{e}$~\cite{DvoStu02},
\begin{equation}\label{eq:q1q2nue}
  q_{\nu_{e},f}^{(1)} =
  I_{\mathrm{L}3}^{(f)}-2Q_{f}\sin^{2}\theta_{\mathrm{W}}+\delta_{ef},
  \quad
  q_{\nu_{e},f}^{(2)} =
  -I_{\mathrm{L}3}^{(f)}-\delta_{ef},
\end{equation}
where $I_{\mathrm{L}3}^{(f)}$ is the third component of the weak
isospin of type $f$ fermions, $Q_{f}$ is the value of their electric
charge, $\theta_{\mathrm{W}}$ is the Weinberg angle, $\delta_{ef}=1$
for electrons and vanishes for protons and neutrons. To get these
coefficients for $\nu_{\mu,\tau}$ the symbol $\delta_{ef}$ should
be eliminated from eq.~(\ref{eq:q1q2nue}).

It was experimentally confirmed (see, e.g., ref.~\cite{QiaWan14})
that the flavor neutrino eigenstates are the superposition of the
neutrino mass eigenstates, $\psi_{i}$, $i=1,2,\dotsc$,
\begin{equation}\label{eq:nupsi}
  \nu_{\alpha}=\sum_{i}U_{\alpha i}\psi_{i},
\end{equation}
where $\left(U_{\alpha i}\right)$ is the unitary mixing matrix. The
transformation in eq.~(\ref{eq:nupsi}) diagonalizes the neutrino
mass matrix. Only using the neutrino mass eigenstates we can reveal
the nature of neutrinos, i.e. say whether they are Dirac or Majorana
particles. Despite the great experimental efforts to shed light upon the nature of neutrinos~\cite{Pet13}, this issue still remains open. Here we shall suppose that $\psi_{i}$ correspond to Dirac fields.

The effective Lagrangian for the interaction of $\psi_{i}$ with background
matter can be obtained using eqs.~(\ref{eq:Lf}) and~(\ref{eq:nupsi}),
\begin{equation}\label{eq:Lg}
  \mathcal{L}_{\mathrm{eff}} =
  - \sum_{ij}
  \bar{\psi}_{i}\gamma_{\mu}^{\mathrm{L}}\psi_{j}\cdot g_{ij}^{\mu},
\end{equation}
where
\begin{equation}\label{eq:gab}
  g_{ij}^{\mu} = \sum_{\alpha}U_{\alpha i}^{*}U_{\alpha j}f_{\alpha}^{\mu},
\end{equation}
is the nondiagonal effective potential in the mass eigenstates basis.

Using eq.~(\ref{eq:Lg}) one obtains that the corresponding Dirac
equations for the neutrino mass eigenstates are coupled,
\begin{equation}\label{eq:Depsi}
  \left[
    \mathrm{i}\gamma^{\mu}\partial_{\mu}-m_{i}-\gamma_{\mu}^{\mathrm{L}}g_{ii}^{\mu}
  \right]
  \psi_{i} =
  \sum_{j\neq i}\gamma_{\mu}^{\mathrm{L}}g_{ij}^{\mu}\psi_{j},
\end{equation}
where $m_{i}$ is the mass of $\psi_{i}$. One can proceed in the
analytical analysis of eq.~(\ref{eq:Depsi}) if we exactly account
for only the diagonal effective potentials $g_{ii}^{\mu}$. To take
into account the r.h.s. of eq.~(\ref{eq:Depsi}), depending on the
nondiagonal elements of the matrix $\left(g_{ij}^{\mu}\right)$, with
$i\neq j$, one should apply a perturbative method (see section~\ref{sec:OSC}
below).

\section{Massive neutrinos in noninertial frames\label{sec:MASSNUNONIF}}

In this section we generalize the Dirac equation for a neutrino interacting
with a background matter to the situation when the velocity of the
matter motion is not constant. In particular, we study the case of
the matter rotation with a constant angular velocity. Then we obtain the
solution of the Dirac equation and find the energy spectrum.

If we discuss a neutrino mass eigenstate propagating in a nonuniformly
moving matter, the expressions for $f_{\alpha}^{\mu}$ in eqs.~(\ref{eq:flavnueffp})-(\ref{eq:lambdams})
become invalid since they are derived under the assumption of the
unbroken Lorentz invariance. The most straightforward way to describe
the neutrino evolution in matter moving with an acceleration is to
rewrite the Dirac equation for a neutrino in the noninertial frame
where matter is at rest. In this case one can unambiguously define
the components of $f_{\alpha}^{\mu}$. Assuming that background fermions
are unpolarized, we find that in this reference frame
\begin{equation}\label{eq:f0nonin}
  f_{\alpha}^{0}=\sqrt{2}G_{\mathrm{F}}\sum_{f}q_{\alpha,f}^{(1)}n_{f}\neq0,
\end{equation}
with the rest of the effective potentials being equal to zero.

It is known that the motion of a test particle in a noninertial frame
is equivalent to the interaction of this particle with a gravitational
field. The Dirac equation for a massive neutrino moving in a curved space-time
and interacting with background matter can be obtained by the generalization
of eq.~(\ref{eq:Depsi}) (see also ref.~\cite{GriMamMos80}),
\begin{equation}\label{eq:Depsicurv}
  \left[
    \mathrm{i}\gamma^{\mu}(x)\nabla_{\mu}-m
  \right]
  \psi =
  \frac{1}{2}\gamma_{\mu}(x)g^{\mu}
  \left[
    1-\gamma^{5}(x)
  \right]
  \psi,
\end{equation}
where $\gamma_{\mu}(x)$ are the coordinate dependent Dirac matrices,
$\nabla_{\mu}=\partial_{\mu}+\Gamma_{\mu}$ is the covariant derivative,
$\Gamma_{\mu}$ is the spin connection, $\gamma^{5}(x) = -\tfrac{\mathrm{i}}{4!} E^{\mu\nu\alpha\beta} \gamma_{\mu}(x) \gamma_{\nu}(x) \gamma_{\alpha}(x) \gamma_{\beta}(x)$,
$E^{\mu\nu\alpha\beta} = \tfrac{1}{\sqrt{-g}} \varepsilon^{\mu\nu\alpha\beta}$
is the covariant antisymmetric tensor in curved space-time, and $g=\det(g_{\mu\nu})$ is the determinant of the metric tensor $g_{\mu\nu}$. Note that in eq.~(\ref{eq:Depsicurv})
we account for only the diagonal neutrino interaction with matter.
That is why we omit the index $i$ in order not to encumber the notation:
$m\equiv m_{i}$ etc. It should be noted that analogous Dirac equation
was discussed in ref.~\cite{PirRoyWud96}.

We shall be interested in the neutrino motion in matter rotating with
the constant angular velocity $\omega$. Choosing the corotating frame
we get that only $g^{0} \equiv g^{0}_{ii}$ is nonvanishing, cf. eqs.~(\ref{eq:gab})
and~(\ref{eq:f0nonin}).

\subsection{Neutrino motion in a rotating frame\label{sub:ROTATION}}

The interval in the rotating frame is~\cite{LanLif94}
\begin{equation}\label{eq:mertrot}
  \mathrm{d}s^{2} =
  g_{\mu\nu}\mathrm{d}x^{\mu}\mathrm{d}x^{\nu} =
  (1-\omega^{2}r^{2})\mathrm{d}t^{2} -
  \mathrm{d}r^{2}-2\omega r^{2}\mathrm{d}t\mathrm{d}\phi -
  r^{2}\mathrm{d}\phi^{2}-\mathrm{d}z^{2},
\end{equation}
where we use the cylindrical coordinates $x^{\mu}=(t,r,\phi,z)$.%
One can check that the metric tensor in eq.~(\ref{eq:mertrot})
can be diagonalized, $\eta_{ab}=e_{a}^{\ \mu}e_{b}^{\ \nu}g_{\mu\nu}$,
if use the following vierbein vectors:
\begin{align}\label{eq:vierb}
  e_{0}^{\ \mu}= &
  \left(
    \frac{1}{\sqrt{1-\omega^{2}r^{2}}},0,0,0
  \right),
  \nonumber
  \\
  e_{1}^{\ \mu}= & (0,1,0,0),
  \nonumber
  \\
  e_{2}^{\ \mu}= &
  \left(
    \frac{\omega r}{\sqrt{1-\omega^{2}r^{2}}},0,\frac{\sqrt{1-\omega^{2}r^{2}}}{r},0
  \right),
  \nonumber
  \\
  e_{3}^{\ \mu}= & (0,0,0,1).
\end{align}
Here $\eta_{ab}=\text{diag}(1,-1,-1,-1)$ is the metric in a locally
Minkowskian frame.

Let us introduce the Dirac matrices in a locally Minkowskian frame
by $\gamma^{\bar{a}}=e_{\ \mu}^{a}\gamma^{\mu}(x)$, where $e_{\ \mu}^{a}$
is the inverse vierbein: $e_{\ \mu}^{a}e_{b}^{\ \mu}=\delta_{b}^{a}$.
Starting from now, we shall mark an index with a bar to show that
a gamma matrix is defined in a locally Minkowskian frame. We notice
that $g=-\left[\det(e_{\ \mu}^{a})\right]^{2}$ and $E^{\mu\nu\alpha\beta} e_{\ \mu}^{m} e_{\ \nu}^{n} e_{\ a}^{\alpha} e_{\ b}^{\beta} = \varepsilon^{mnab}$.
Thus $\gamma^{5}(x) = \mathrm{i} \gamma^{\bar{0}} \gamma^{\bar{1}} \gamma^{\bar{2}} \gamma^{\bar{3}} = \gamma^{\bar{5}}$
does not depend on coordinates.

The spin connection has the form~\cite{GriMamMos80}
\begin{equation}
  \Gamma_{\mu} = -\frac{\mathrm{i}}{4}\sigma^{ab}\omega_{ab\mu},
  \quad
  \omega_{ab\mu}=e_{a}^{\ \nu}e_{b\nu;\mu},
\end{equation}
where $\sigma_{ab}=\tfrac{\mathrm{i}}{2}[\gamma_{\bar{a}},\gamma_{\bar{b}}]_{-}$
are the generators of the Lorentz transformations in a locally Minkowskian
frame. Here the semicolon stays for the covariant derivative: $A_{\nu;\mu}=\partial_{\mu}A_{\nu}-\Gamma_{\ \mu\nu}^{\alpha}A_{\alpha}$,
with $\Gamma_{\ \mu\nu}^{\alpha}$ being the Christoffel symbol of
the second kind. The components of the connection one-form $\omega_{ab}=\omega_{ab\mu}\mathrm{d}x^{\mu}$
are
\begin{align}\label{eq:spinconn}
  \omega_{01\mu}= & -\omega_{10\mu}=
  \left(
    -\frac{\omega^{2}r}{\sqrt{1-\omega^{2}r^{2}}}, 0,
    -\frac{\omega r}{\sqrt{1-\omega^{2}r^{2}}}, 0
  \right),
  \nonumber
  \\
  \omega_{02\mu}= & -\omega_{20\mu}=
  \left(
    0,\frac{\omega}{1-\omega^{2}r^{2}},0,0
  \right),
  \nonumber
  \\
  \omega_{12\mu}= & -\omega_{21\mu} =
  \left(
    \frac{\omega}{\sqrt{1-\omega^{2}r^{2}}},0,\frac{1}{\sqrt{1-\omega^{2}r^{2}}},0
  \right),
\end{align}
with the rest of the components being equal to zero.%

Using eqs.~(\ref{eq:vierb})-(\ref{eq:spinconn}), the Dirac eq.~(\ref{eq:Depsicurv})
can be rewritten as
\begin{align}\label{eq:Derotf}
  [\mathcal{D} - & m]\psi =
  \frac{1}{2} \sqrt{1-\omega^{2}r^{2}}\gamma^{\bar{0}}g^{0}
  (1-\gamma^{\bar{5}})\psi,
  \nonumber
  \\
  \mathcal{D} = &
  \mathrm{i} \frac{\gamma^{\bar{0}}+\omega r\gamma^{\bar{2}}}{\sqrt{1-\omega^{2}r^{2}}}
  \partial_{0} +
  \mathrm{i} \gamma^{\bar{1}}
  \left(
    \partial_{r}+\frac{1}{2r}
  \right) +
  \mathrm{i} \gamma^{\bar{2}}
  \frac{\sqrt{1-\omega^{2}r^{2}}}{r}\partial_{\phi} +
  \mathrm{i} \gamma^{\bar{3}}\partial_{z}
  \notag
  \\
  & -
  \frac{\omega}{2(1-\omega^{2}r^{2})}
  \gamma^{\bar{3}}\gamma^{\bar{5}}.
\end{align}
The analogous Dirac equation was recently derived in ref.~\cite{Bak13}.
Since eq.~(\ref{eq:Derotf}) does not explicitly contain $t$, $\phi$,
and $z$, its solution can be expressed as
\begin{equation}\label{eq:psitildepsi}
  \psi =
  \exp
  \left(
    -\mathrm{i}Et+\mathrm{i}J_{z}\phi+\mathrm{i}p_{z}z
  \right)
  \psi_{r},
\end{equation}
where $\psi_{r}=\psi_{r}(r)$ is the spinor depending on the radial
coordinate, $J_{z}=\tfrac{1}{2}-l$ (see, e.g., ref.~\cite{SchWieGre83}),
and $l=0,\pm1,\pm2,\dotsc$.

In eq.~(\ref{eq:Derotf}) one can neglect terms $\sim(\omega r)^{2}$.
Indeed, if we study a neutrino in a rotating pulsar, then $r\lesssim10\thinspace\text{km}$
and $\omega\lesssim10^{3}\thinspace\text{s}^{-1}$. Thus $(\omega r)^{2}\lesssim1.1\times10^{-3}$
is a small parameter. Therefore eq.~(\ref{eq:Derotf}) can be transformed
to
\begin{multline}\label{eq:Depsisimp}
  \bigg[
    \mathrm{i}\gamma^{\bar{1}}
    \left(
      \partial_{r}+\frac{1}{2r}
    \right) -
    \gamma^{\bar{2}}
    \left(
      \frac{J_{z}}{r}-\omega r E
    \right) +
    \gamma^{\bar{0}}
    \left(
      E-\frac{g^{0}}{2}
    \right) -
    \gamma^{\bar{3}}p_{z}
    \\
    + \frac{g^{0}}{2}\gamma^{\bar{0}}\gamma^{\bar{5}} -
    \frac{\omega}{2}\gamma^{\bar{3}}\gamma^{\bar{5}} - m
  \bigg]
  \psi_{r}=0,
\end{multline}
where we keep only the terms linear in $\omega$. It should be noted
that the term $\sim\omega\gamma^{\bar{3}}\gamma^{\bar{5}}$ in eq.~(\ref{eq:Depsisimp})
is equivalent to the neutrino interaction with matter moving with
an effective velocity.

Let us transform the wave function as
\begin{equation}\label{eq:boost3}
  \psi_{r} = U_{3}\tilde{\psi}_{r},
  \quad
  U_{3} = \exp
  \left(
    \frac{\upsilon}{2}\gamma^{\bar{0}}\gamma^{\bar{3}}
  \right) =
  \cosh\frac{\upsilon}{2} +
  \gamma^{\bar{0}}\gamma^{\bar{3}} \sinh\frac{\upsilon}{2},
\end{equation}
where $\upsilon$ is a real parameter satisfying $\tanh \upsilon =\omega/g^{0}$.
Such a transformation is equivalent to a boost. If we again study
a neutrino moving inside a rotating pulsar, then $g^{0}\sim G_{\mathrm{F}}n_{n}\sim10\thinspace\text{eV}$,
where $n_{n}\sim10^{38}\thinspace\text{cm}^{-3}$ is the neutron density.
Thus $\upsilon\sim10^{-13}\ll1$ and the transformation in eq.~(\ref{eq:boost3})
exists. The spinor $\tilde{\psi}_{r}$ obeys the equation,
\begin{equation}\label{eq:Detildepsicov}
  \left[
    \gamma^{\bar{a}}Q_{a} +
    \frac{\tilde{g}^{0}}{2}\gamma^{\bar{0}}\gamma^{\bar{5}} - m
  \right]\tilde{\psi}_{r}=0,
\end{equation}
where $Q^{a}= \left( \tilde{E},-\mathrm{i} \left( \partial_{r}+\tfrac{1}{2r} \right) , \left( \frac{J_{z}}{r}-\omega rE \right) , \tilde{p}_{z} \right)$.
It is convenient to represent $Q^{a}=q^{a}-q_{\mathrm{eff}}A_{\mathrm{eff}}^{a}$,
where $q^{a} = \left( \tilde{E},-\mathrm{i}\partial_{r},0,p_{z} \right)$,
$A_{\mathrm{eff}}^{a} = \left( 0,\tfrac{\mathrm{i}}{2q_{\mathrm{eff}}r},\tfrac{1}{q_{\mathrm{eff}}} \left( \omega rE-\frac{J_{z}}{r} \right) , 0\right)$
is the vector potential of the effective electromagnetic field, and
$q_{\mathrm{eff}}$ is the effective electric charge. Here
\begin{align}
  \tilde{E}= & E' \cosh \upsilon - p_{z} \sinh \upsilon,
  \nonumber
  \\
  \tilde{p}_{z}= & p_{z} \cosh \upsilon - E' \sinh \upsilon,
  \nonumber
  \\
  \tilde{g}^{0}= & g^{0} \cosh \upsilon - \omega \sinh \upsilon,
\end{align}
and $E'=E-\tfrac{g^{0}}{2}$.

We shall look for the solution of eq.~(\ref{eq:Detildepsicov}) as
\begin{equation}\label{eq:tildepsiphi}
  \tilde{\psi}_{r}=\Pi\Phi,
  \quad
  \Pi=\gamma^{\bar{a}}Q_{a}-\frac{\tilde{g}^{0}}{2}\gamma^{\bar{0}}\gamma^{\bar{5}} + m,
\end{equation}
where $\Phi=\Phi(r)$ is a new spinor. This spinor obeys the equation,
\begin{multline}\label{eq:quadeqphigen}
  \bigg[
    \partial_{r}^{2} + \frac{\partial_{r}}{r} - \frac{1}{4r^2} +
    \tilde{E}^{2} - \tilde{p}_{z}^{2} -
    \left(
      \frac{J_{z}}{r} - \omega rE
    \right)^{2} -
    m^{2} +
    \frac{
    \left(
      \tilde{g}^{0}
    \right)^{2}}{4}
    \\
    +
    \left(
      E\omega+\frac{J_{z}}{r^{2}}
    \right)
    \Sigma_{3} - \tilde{g}^{0}\tilde{E}\gamma^{\bar{5}} +
    m \tilde{g}^{0}\gamma^{\bar{0}}\gamma^{\bar{5}}
  \bigg]
  \Phi=0,
\end{multline}
which can be obtained using eq.~(\ref{eq:Detildepsicov}).

Equation~(\ref{eq:quadeqphigen}) can be solved for ultrarelativistic
neutrinos when the particle mass is neglected. In this case we
represent $\Phi=v\varphi$, where $v$ is a constant spinor and $\varphi=\varphi(r)$
is a scalar function. We notice that both $\Sigma_{3}=\gamma^{\bar{0}}\gamma^{\bar{3}}\gamma^{\bar{5}}$
and $\gamma^{\bar{5}}$ commute with the operator of eq.~(\ref{eq:quadeqphigen})
when $m\to0$. Taking into account that $\left[ \Sigma_{3},\gamma^{\bar{5}} \right]_{-} = 0$,
one gets that the spinor $v$ satisfies $\Sigma_{3}v=\sigma v$ and
$\gamma^{\bar{5}}v=\chi v$, where $\sigma=\pm1$ and $\chi=\pm1$.

We can choose two linearly independent spinors $v$ as
\begin{equation}\label{eq:vspinors}
  v_{-}=v(\chi=+1,\sigma=-1) =
  \left(
    \begin{array}{c}
      0 \\
      1 \\
      0 \\
      0
    \end{array}
  \right),
  \quad
  v_{+}=v(\chi=-1,\sigma=+1) =
  \left(
    \begin{array}{c}
      0 \\
      0 \\
      1 \\
      0
    \end{array}
  \right).
\end{equation}
To derive eq.~(\ref{eq:vspinors}) we use the Dirac matrices in form,
\begin{equation}\label{eq:gammamatr}
  \gamma^{\bar{0}} =
  \left(
    \begin{array}{cc}
      0 & -1 \\
      -1 & 0
    \end{array}
  \right),
  \quad
  \gamma^{\bar{k}} =
  \left(
    \begin{array}{cc}
      0 & \sigma_{k} \\
      -\sigma_{k} & 0
    \end{array}
  \right),
  \quad
  \gamma^{\bar{5}} =
  \left(
    \begin{array}{cc}
      1 & 0 \\
      0 & -1
    \end{array}
  \right),
\end{equation}
that corresponds to the chiral representation~\cite{ItzZub80}. In
eq.~(\ref{eq:gammamatr}) $\sigma_{k}$ are the Pauli matrices.

Introducing the new variable $\rho=E\omega r^{2}$, we get the equation
for $\varphi_{\sigma}$,
\begin{equation}\label{eq:phisigma}
  \left[
    \rho\frac{\mathrm{d}^{2}}{\mathrm{d}\rho} +
    \frac{\mathrm{d}}{\mathrm{d}\rho} -
    \frac{1}{4\rho}
    \left(
      l+\frac{\sigma-1}{2}
    \right)^{2} -
    \frac{\rho}{4} -
    \frac{1}{2}
    \left(
      l-\frac{\sigma+1}{2}
    \right) +
    \kappa
  \right]
  \varphi_{\sigma}=0,
\end{equation}
where
\begin{equation}\label{eq:kappa}
  \kappa =
  \frac{1}{4E\omega}
  \left(
    \tilde{E}^{2}-\tilde{p}_{z}^{2} +
    \frac{
    \left(
      \tilde{g}^{0}
    \right)^{2}}{4} -
    \tilde{g}^{0}\tilde{E}\chi
  \right).
\end{equation}
The solution of eq.~(\ref{eq:phisigma}) which vanishes at the infinity,
$\varphi_{\sigma}(\rho\to\infty)\to0$, can be expressed as $\varphi_{+}=I_{N,s}$
and $\varphi_{-}=I_{N-1,s}$, where $N=0,1,2,\dotsc$, $s=N-l$, and
$I_{N,s}=I_{N,s}(\rho)$ is the Laguerre function,
\begin{equation}\label{eq:INs}
  I_{N,s}(\rho) =
  \sqrt{\frac{s!}{N!}}\exp
  \left(
    -\frac{\rho}{2}
  \right)\rho^{\frac{N-s}{2}}L_{s}^{N-s}(\rho).
\end{equation}
Here $L_{s}^{\alpha}(\rho)$ are the associated Laguerre polynomials.

The energy spectrum can be obtained if we notice that $\kappa=N$
in eq.~(\ref{eq:kappa}). Thus we get
\begin{align}\label{eq:Energylev}
  \left[
    E_{A}-2N\omega-g^{0}
  \right]^{2} =
  & (2N\omega)^{2}+4N\omega g^{0} +
  \left(
    p_{z}-\frac{\omega}{2}
  \right)^{2},
  \nonumber
  \\
  \left[
    E_{S}-2N\omega
  \right]^{2} =
  & (2N\omega)^{2} +
  \left(
    p_{z}+\frac{\omega}{2}
  \right)^{2},
\end{align}
where $E_{A}=E(\chi=+1)$ and $E_{S}=E(\chi=-1)$ are the energies
of active and sterile neutrinos respectively. Comparing the expression
for $E_{S}\approx2N\omega+\sqrt{(2N\omega)^{2}+p_{z}^{2}}$ with the
energy of a neutrino in an inertial nonrotating frame $\sqrt{\mathbf{p}_{\perp}^{2}+p_{z}^{2}}$,
where $\mathbf{p}_{\perp}$ is the momentum in the equatorial plane,
we can identify $2N\omega$ inside the square root as $|\mathbf{p}_{\perp}|$.
It should be also noted that the term $2N\omega$, which additively
enters to both $E_{A}$ and $E_{S}$, is due to the noninertial effects
for a Dirac fermion in a rotating frame~\cite{HehNi90}.

Using eqs.~(\ref{eq:boost3}), (\ref{eq:tildepsiphi}), (\ref{eq:vspinors}),
and~(\ref{eq:Energylev}) we can get the wave function $\psi_{r}$
in the explicit form,
\begin{equation}\label{eq:psiLR}
  \psi_{r}^{\mathrm{L}} =
  U_{3}\Pi v_{-}\varphi_{-}=
  \left(
    \begin{array}{c}
      0 \\
      \eta
    \end{array}
  \right),
  \quad
  \psi_{r}^{\mathrm{R}} =
  U_{3}\Pi v_{+}\varphi_{+} =
  \left(
    \begin{array}{c}
      \xi \\
      0
    \end{array}
  \right),
\end{equation}
where
\begin{equation}\label{eq:etaxi}
  \eta =
  \left(
    \begin{array}{c}
      -\mathrm{i}C_{1}I_{N,s} \\
      C_{2}I_{N-1,s}
    \end{array}
  \right),
  \quad
  \xi =
  \left(
    \begin{array}{c}
      C_{3}I_{N,s} \\
      -\mathrm{i}C_{4}I_{N-1,s}
    \end{array}
  \right),
\end{equation}
and
\begin{align}\label{eq:C1-4gen}
  C_{1}^{2}= &
  \frac{E_{A}\omega}{\pi}
  \frac{
  \left(
    1-\omega/g^{0}
  \right)^{3/2}}
  {
  \left(
    1+\omega/g^{0}
  \right)^{1/2}}
  \frac{4N\omega E_{A}}{
  \left(
    \tilde{E}_{A}+\tilde{p}_{z}-\tilde{g}^{0}/2
  \right)^{2}
  \left(
    1+\omega/g^{0}
  \right) +
  4N\omega E_{A}
  \left(
    1-\omega/g^{0}
  \right)},
  \nonumber
  \\
  C_{2}^{2}= &
  \frac{E_{A}\omega}{\pi}
  \frac{
  \left(
    1+\omega/g^{0}
  \right)^{3/2}}
  {
  \left(
    1-\omega/g^{0}
  \right)^{1/2}}
  \frac{
  \left(
    \tilde{E}_{A}+\tilde{p}_{z}-\tilde{g}^{0}/2
  \right)^{2}}
  {
  \left(
    \tilde{E}_{A}+\tilde{p}_{z}-\tilde{g}^{0}/2
  \right)^{2}
  \left(
    1+\omega/g^{0}
  \right) +
  4N\omega E_{A}
  \left(
    1-\omega/g^{0}
  \right)},
  \nonumber
  \\
  C_{3}^{2}= &
  \frac{E_{S}\omega}{\pi}
  \frac{
  \left(
    1+\omega/g^{0}
  \right)^{3/2}}
  {
  \left(
    1-\omega/g^{0}
  \right)^{1/2}}
  \frac{
  \left(
    \tilde{E}_{S}+\tilde{p}_{z}+\tilde{g}^{0}/2
  \right)^{2}}
  {
  \left(
    \tilde{E}_{S}+\tilde{p}_{z}+\tilde{g}^{0}/2
  \right)^{2}
  \left(
    1+\omega/g^{0}
  \right) +
  4N\omega E_{S}
  \left(
    1-\omega/g^{0}
  \right)},
  \nonumber
  \\
  C_{4}^{2}= &
  \frac{E_{S}\omega}{\pi}
  \frac{
  \left(
    1-\omega/g^{0}
  \right)^{3/2}}
  {
  \left(
    1+\omega/g^{0}
  \right)^{1/2}
  }
  \frac{4N\omega E_{S}}{
  \left(
    \tilde{E}_{S}+\tilde{p}_{z}+\tilde{g}^{0}/2
  \right)^{2}
  \left(
    1+\omega/g^{0}
  \right) +
  4N\omega E_{S}
  \left(
    1-\omega/g^{0}
  \right)
  }.
\end{align}
To derive eqs.~(\ref{eq:psiLR})-(\ref{eq:C1-4gen}) we use the following
properties of the Laguerre functions:
\begin{align}
  \left(
    \partial_{r}-\frac{l-1}{r}-\omega Er
  \right)
  I_{N-1,s}(\omega Er^{2})= &
  - 2\sqrt{N\omega E}I_{N,s}(\omega Er^{2}),
  \nonumber
  \\
  \left(
    \partial_{r}+\frac{l}{r}+\omega Er
  \right)
  I_{N,s}(\omega Er^{2})= &
  2\sqrt{N\omega E}I_{N-1,s}(\omega Er^{2}),
\end{align}
which can be verified using eq.~(\ref{eq:INs}).

In the following we shall study the important case when $\omega\ll g^{0}$.
In this situation we get for the coefficients $C_{i}$,

\begin{align}\label{eq:C1-4slowrot}
  C_{1}^{2} \approx &
  \frac{E_{A}\omega}{2\pi}\frac{E_{A}-p_{z}-g^{0}}{E_{A}-g^{0}},
  \quad
  C_{3}^{2} \approx \frac{\omega}{2\pi}
  \left(
    E_{S}+p_{z}
  \right),
  \nonumber
  \\
  C_{2}^{2} \approx &
  \frac{E_{A}\omega}{2\pi}\frac{E_{A}+p_{z}-g^{0}}{E_{A}-g^{0}},
  \quad
  C_{4}^{2} \approx \frac{\omega}{2\pi}
  \left(
    E_{S}-p_{z}
  \right),
\end{align}
which result from eq.~(\ref{eq:C1-4gen}).

The solution given in eqs.~(\ref{eq:etaxi})-(\ref{eq:C1-4slowrot})
is valid for $N>0$. If $N=0$, the spinors $\eta$ and $\xi$ take
the form
\begin{align}\label{eq:etaxi0}
  \eta= &
  \mathcal{N}_{A}
  \left(
    \begin{array}{c}
      I_{0,0} \\
      0
    \end{array}
  \right),
  \quad
  \xi =
  \mathcal{N}_{S}
  \left(
    \begin{array}{c}
      I_{0,0} \\
      0
    \end{array}
  \right),
  \nonumber
  \\
  \mathcal{N}_{A}= &
  \sqrt{\frac{\omega}{\pi}}
  \left|
    g^{0}+
    \left|
      p_{z}-\frac{\omega}{2}
    \right|
  \right|^{1/2}
  \left(
    \frac{1+\omega/g^{0}}{1-\omega/g^{0}}
  \right)^{1/4}
  \approx
  \sqrt{\frac{\omega}{\pi}
  \left|
    g^{0}+
    \left|
      p_{z}
    \right|
  \right|
  },
  \nonumber
  \\
  \mathcal{N}_{S}= &
  \sqrt{\frac{\omega}{\pi}}
  \left|
    p_{z}+\frac{\omega}{2}
  \right|^{1/2}
  \left(
    \frac{1-\omega/g^{0}}{1+\omega/g^{0}}
  \right)^{1/4}
  \approx
  \sqrt{\frac{\omega}{\pi}
  \left|
    p_{z}
  \right|},
\end{align}
where we also assume that $\omega\ll g^{0}$.

It should be noted that the solutions presented in eqs.~(\ref{eq:psiLR})-(\ref{eq:etaxi0})
satisfy the normalization condition,
\begin{equation}\label{eq:normgen}
  \int\psi_{N,s,p_{z}}^{\dagger}(x)
  \psi_{N',s',p'_{z}}(x)\sqrt{-g}\mathrm{d}^{3}x =
  \delta_{NN'}\delta_{ss'}
  \delta
  \left(
    p_{z}-p'_{z}
  \right).
\end{equation}
Here $\psi$ and $\psi_{r}$ are related by eq.~(\ref{eq:psitildepsi}).

Using eq.~(\ref{eq:quadeqphigen}) we can get the corrections to
the energy levels due to the nonzero mass, $E_{A,S}\to E_{A,S}+E_{A,S}^{(1)}$.
On the basis of eqs.~(\ref{eq:psiLR})-(\ref{eq:C1-4slowrot}) one finds the
expression for $E_{A,S}^{(1)}$ in the limit $\omega\ll g^{0}$,
\begin{equation}\label{eq:EAS1}
  E_{A}^{(1)} =
  \frac{m^{2}}{2
  \left(
    E_{A}-2N\omega-g^{0}
  \right)
  },
  \quad
  E_{S}^{(1)} =
  \frac{m^{2}}{2
  \left(
    E_{S}-2N\omega
  \right)
  },
\end{equation}
Note that eq.~(\ref{eq:EAS1}) can be obtained if we retain $m^{2}$
in the operator of eq.~(\ref{eq:quadeqphigen}).

If we discuss neutrinos moving along the rotation axis, then $2N\omega\ll|p_{z}|$.
Using eq.~(\ref{eq:Energylev}) we get the energy levels of active
neutrinos in this case
\begin{equation}\label{eq:Eadecom}
  E_{A} =
  |p_{z}|+g^{0}
  \left(
    1+\frac{2N\omega}{|p_{z}|}
  \right) +
  2N\omega +
  \frac{2(N\omega)^{2}}{|p_{z}|} +
  \frac{m^{2}}{2|p_{z}|},
\end{equation}
where we also keep the mass correction in eq.~(\ref{eq:EAS1}). One
can see in eq.~(\ref{eq:Eadecom}) that $|p_{z}|+g^{0}+\frac{m^{2}}{2|p_{z}|}$
corresponds to the energy of a left-handed neutrino
interacting with background matter in a flat space-time. The rest
of the terms in eq.~(\ref{eq:Eadecom}) are the corrections due to
the matter rotation.


\section{Flavor oscillations of Dirac neutrinos in rotating matter\label{sec:OSC}}

In this section we study the evolution of the system of massive mixed
neutrinos in rotating matter. We formulate the initial condition for
this system and derive the effective Schr\"{o}dinger equation which governs
neutrino flavor oscillations. Then we find the correction to the resonance
condition owing to the matter rotation and estimate its value for
a millisecond pulsar.

We can generalize the results of section~\ref{sec:MASSNUNONIF} to include
different neutrino eigenstates. The interaction of neutrino mass eigenstates
with background matter is nondiagonal, cf. eq.~(\ref{eq:Lg}). Therefore
the generalization of eq.~(\ref{eq:Derotf}) for several mass eigenstates
$\psi_{i}$ reads
\begin{equation}\label{eq:Derotfpsia}
  \left[
    \mathcal{D}-m_{i}
  \right] \psi_{i} =
  \frac{1}{2}\gamma^{\bar{0}}
  g_{i}^{0} (1-\gamma^{\bar{5}}) \psi_{i} +
  \frac{1}{2}\gamma^{\bar{0}}
  \sum_{i\neq j}
  g_{ij}^{0}(1-\gamma^{\bar{5}})\psi_{j},
\end{equation}
where $g_{i}^{0}\equiv g_{ii}^{0}$ and $g_{ij}^{0}$ are the time
components of the matrix $\left(g_{ij}^{\mu}\right)$ given in eq.~(\ref{eq:gab}),
$m_{i}$ is the mass of $\psi_{i}$, and $\mathcal{D}$ can be found
in eq.~(\ref{eq:Derotf}). As in section~\ref{sec:MASSNUNONIF}, we
omitted the term $(\omega r)^{2}\ll1$ in eq.~(\ref{eq:Derotfpsia}).
Note that eq.~(\ref{eq:Derotfpsia}) is a generalization of eq.~(\ref{eq:Depsi})
for a system of the neutrino mass eigenstates moving in a rotating
frame.

We shall study the evolution of active ultrarelativistic neutrinos
and neglect neutrino-antineutrino transitions. In this case we can
restrict ourselves to the analysis of two component spinors. The general solution
of eq.~(\ref{eq:Derotfpsia}) has the form,
\begin{equation}\label{eq:gensolosc}
  \eta_{i}(x) =
  \sum_{N,s}
  \int\frac{\mathrm{d}p_{z}}{\sqrt{2\pi}}
  a_{N,s,p_{z}}^{(i)}
  e^{\mathrm{i}p_{z}z+\mathrm{i}J_{z}\phi}
  u_{N,s,p_{z}}(r)
  e^{-\mathrm{i}E_{i}t},
\end{equation}
where
\begin{equation}\label{eq:upm}
  u_{N,s,p_{z}} = \sqrt{\frac{|p_{z}|\omega}{\pi}}
  \begin{cases}
    \left(
      \begin{array}{c}
        0 \\
        I_{N-1,s}(\omega|p_{z}|r^{2})
      \end{array}
    \right), & \text{if}\ p_{z}>0,
    \\
    \left(
      \begin{array}{c}
        -\mathrm{i}I_{N,s}(\omega|p_{z}|r^{2}) \\
        0
      \end{array}
    \right), & \text{if}\ p_{z}<0.
  \end{cases}
\end{equation}
The energy levels $E_{i}$ are given in eq.~(\ref{eq:Eadecom}) with
$m\to m_{i}$. Here we omit the subscript $A$ in order not to encumber
the notation. Our goal is to find the coefficient $a_{N,s,p_{z}}^{(i)}=a_{N,s,p_{z}}^{(i)}(t)$.
In eqs.~(\ref{eq:gensolosc}) and~(\ref{eq:upm}) we neglect the
small ratio $\omega/g_{i}^{0}$.

Although in general case the dynamics of a Dirac field obeys the quantum
field theory, is some situations, related to the description of neutrino
oscillations, we can use first quantized spinors (for the detailed
analysis see ref.~\cite{Dvo11} and references therein). Thus we
will suppose that the quantity $a_{N,s,p_{z}}^{(i)}$ in eq.~(\ref{eq:gensolosc})
is a $c$-number coefficient rather than an operator.

In the following we shall discuss the case of two neutrino mass eigenstates,
i.e. $i=1,2$, interacting with rotating background matter. In this
situation the mixing matrix $\left(U_{\alpha i}\right)$ in eq.~(\ref{eq:nupsi})
can be parametrized with help of one vacuum mixing angle $\theta$.

To describe the time evolution of the system we should specify the
initial condition. We shall fix the initial wave functions in the
flavor eigenstates basis. One of the possible choices of the initial
wave functions is the following: $\nu_{\alpha}(\mathbf{r},t=0)=0$
and $\nu_{\beta}(\mathbf{r},t=0) \sim \exp\left(\mathrm{i}p_{z}^{(0)}z+\mathrm{i}J_{z}^{(0)}\phi\right)\delta_{N,N_{0}}$,
where $\alpha\neq\beta$ and $p_{z}^{(0)}$, $J_{z}^{(0)}=1/2-N_{0}+s_{0}$,
$N_{0}$, and $s_{0}$ are the quantum numbers of the initial wave
function. In the following we shall omit the index $0$ in order not
to encumber the notations. We will be interested in the evolution
of $\nu_{\alpha}$ in the subsequent moments of time. Supposing that
$\nu_{\beta}\equiv\nu_{e}$ and $\nu_{\alpha}\equiv\nu_{\mu}$ or
$\nu_{\tau}$, we get that the adopted initial condition corresponds
to a typical situation of oscillations of neutrinos emitted at the initial stages of a pulsar evolution: one
looks for $\nu_{\mu}$ or $\nu_{\tau}$ in a neutrino beam, initially
consisting of $\nu_{e}$ only.

Using the fact that $u_{N,s,p_{z}}^{\dagger}\sigma_{2}u_{N,s,p_{z}}=0$,
where $u_{N,s,p_{z}}$ is given in eq.~(\ref{eq:upm}), on the basis
of eq.~(\ref{eq:Derotfpsia}) we get the effective Schr\"{o}dinger equation
for $\tilde{\Psi}^{\mathrm{T}}=(a_{1},a_{2})$,%
\begin{equation}\label{eq:ScheqtildePsi}
  \mathrm{i}\frac{\mathrm{d}\tilde{\Psi}}{\mathrm{d}t} =
  \left(
    \begin{array}{cc}
      0 & g_{12}^{0}
      \exp
      \left[
        \mathrm{i}
          \left(E_{1}-E_{2}
        \right)
        t
      \right]
      \\
      g_{12}^{0}
      \exp
      \left[
        \mathrm{i}
        \left(
          E_{2}-E_{1}
        \right)
        t
      \right] & 0
    \end{array}
  \right)
  \tilde{\Psi}.
\end{equation}
Here we omitted all the indexes of $a_{i}$ besides $i=1,2$. It is
convenient to introduce the modified effective wave function $\Psi=\mathcal{U}_{3}\tilde{\Psi}$,
where $\mathcal{U}_{3}=\text{diag}\left(e^{\mathrm{i}\Omega t/2},e^{-\mathrm{i}\Omega t/2}\right)$,
$\Omega=E_{1}-E_{2}$. Using eq.~(\ref{eq:ScheqtildePsi}), we get
for $\Psi$
\begin{equation}\label{eq:ScheqPsi}
  \mathrm{i}\frac{\mathrm{d}\Psi}{\mathrm{d}t} =
  \left(
    \begin{array}{cc}
      \Omega/2 & g_{12}^{0} \\
      g_{12}^{0} & -\Omega/2
    \end{array}
  \right)
  \Psi.
\end{equation}
Note that eq.~(\ref{eq:ScheqPsi}) has the form of the effective
Schr\"{o}dinger equation one typically deals with in the study of neutrino
flavor oscillations in background matter.

If the transition probability for $\nu_{\alpha}\leftrightarrow\nu_{\beta}$
is close to one, i.e. $P_{\nu_{\beta}\to\nu_{\alpha}}=\left|\left\langle \nu_{\alpha}(t)|\nu_{\beta}(0)\right\rangle \right|^{2}\approx1$,
flavor oscillations of neutrinos are said to be at resonance. Using
eqs.~(\ref{eq:nupsi}), (\ref{eq:f0nonin}), (\ref{eq:Eadecom}), and~\eqref{eq:ScheqPsi},
the resonance condition can be written as,
\begin{equation}\label{eq:rescondgen}
  \left(
    f_{\alpha}^{0}-f_{\beta}^{0}
  \right)
  \left(
    1+\frac{2N\omega}{|p_{z}|}
  \right) +
  \frac{\Delta m^{2}}{2|p_{z}|}\cos2\theta=0,
\end{equation}
where $\Delta m^{2}=m_{1}^{2}-m_{2}^{2}$ is the mass squared difference.

Let us consider electroneutral background matter composed of electrons,
protons, and neutrons. If we study the $\nu_{e}\to\nu_{\alpha}$ oscillation
channel, where $\alpha=\mu,\tau$, using eq.~(\ref{eq:q1q2nue}),
we get that $f_{\nu_{\alpha}}^{0}=-\tfrac{1}{\sqrt{2}}G_{\mathrm{F}}n_{n}$
and $f_{\nu_{\beta}}^{0}\equiv f_{\nu_{e}}^{0}=\sqrt{2}G_{\mathrm{F}}\left(n_{e}-\tfrac{1}{2}n_{n}\right)$,
where $n_{e}$ and $n_{n}$ are the densities of electrons and neutrons.
Using eq.~(\ref{eq:rescondgen}), we obtain that
\begin{equation}\label{eq:resnuenumu}
  \sqrt{2}G_{\mathrm{F}}n_{e}
  \left(
    1+\frac{2N\omega}{|p_{z}|}
  \right) =
  \frac{\delta m^{2}}{2|p_{z}|}\cos2\theta.
\end{equation}
At the absence of rotation, $\omega=0$, eq.~(\ref{eq:resnuenumu})
is equivalent to the Mikheyev-Smirnov-Wolfenstein resonance condition
in background matter~\cite{BleSmi13}.

Let us evaluate the contribution of the matter rotation to the resonance
condition in eq.~(\ref{eq:resnuenumu}) for a neutrino emitted inside
a rotating pulsar. We make a natural assumption that for a corotating
observer neutrinos are emitted in a spherically symmetric way
from a neutrinosphere. That is we should take that $l\approx0$ and
$N\approx s$. Then the trajectory of a neutrino is deflected because
of the noninertial effects and the interaction with background matter.
The radius $\mathcal{R}$ of the trajectory can be found from
\begin{equation}\label{eq:trajrad}
  \mathcal{R}^{2} =
  2|p_{z}|\omega
  \int_{0}^{\infty}
  r^{2}|u_{N,s,p_{z}}(r)|^{2}r\mathrm{d}r
  \approx
  \frac{2N}{|p_{z}|\omega},
\end{equation}
where we take into account that $N\gg1$.

We shall assume that $\mathcal{R}\sim R_{\mathrm{0}}$, where $R_{\mathrm{0}}=10\thinspace\text{km}$
is the pulsar radius. In this case neutrinos escape a pulsar. Taking
that $\omega=10^{3}\thinspace\text{s}^{-1}$ and using eq.~(\ref{eq:trajrad}),
we get that the correction to the resonance condition in eq.~(\ref{eq:resnuenumu})
is $\frac{2N\omega}{|p_{z}|}\approx\left(R_{\mathrm{0}}\omega\right)^{2}\approx10^{-3}$.
The obtained correction to the effective number density is small but
nonzero. This result corrects our previous statement in ref.~\cite{Dvo11b}
that a matter rotation does not contribute neutrino flavor oscillations.

\section{Summary and discussion\label{sec:SUMMARY}}

In summary we notice that we have studied the evolution of massive
mixed neutrinos in nonuniformly moving background matter. The interaction
of neutrinos with background fermions is described in frames of the
Fermi theory (see section~\ref{sec:NUMATFLS}). A particular case of the
matter rotating with a constant angular velocity has been studied
in section~\ref{sub:ROTATION}. We have derived the Dirac equation for
a weakly interacting neutrino in a rotating frame and found its solution
in case of ultrarelativistic neutrinos, cf. eqs.~(\ref{eq:psiLR})-(\ref{eq:C1-4gen}).
The energy spectrum obtained in eqs.~(\ref{eq:Energylev}) and~(\ref{eq:EAS1})
includes the correction owing to the nonzero neutrino mass.

We have used the Dirac equation in a noninertial frame, cf. eq.~(\ref{eq:Depsicurv}),
as a main tool for the study of the neutrino motion in matter moving
with an acceleration. To develop the quantum mechanical description
of such a neutrino we have chosen a noninertial frame where matter
is at rest. In this frame the effective potential of the neutrino-matter
interaction is well defined. However, the wave equation for a neutrino
turns out to be more complicated since one has to deal with noninertial
effects.

Previously neutrinos weakly interacting with a rotating matter were
considered in ref.~\cite{BalPopStu09}. To describe the neutrino-matter
interaction the authors of ref.~\cite{BalPopStu09} used the effective
potential in eq.~(\ref{eq:flavnueffp}) with a nonzero velocity of
background fermions $\mathbf{u}_{f}$ corresponding to a rotation.
It should be, however, noted that the formalism for the study of the
neutrino interaction with moving matter derived in ref.~\cite{LobStu01}
is valid when matter has a constant velocity. That is why, despite
the similarity of the neutrino wave functions, cf. eqs.~(\ref{eq:etaxi})
and~(28) in ref.~\cite{BalPopStu09} (they are both expressed via
Laguerre functions), there is a difference in the neutrino energy
spectrum, cf. eqs.~(\ref{eq:Energylev}) and~(31) in ref.~\cite{BalPopStu09}.

One can suggest the following explanation of the energy levels discrepancy.
It was assumed in ref.~\cite{BalPopStu09} that a neutrino is initially
created in an inertial frame and then starts to interact with a rotating
matter. This situation can be implemented if one studies a neutrino
emission in a very center of a rotating pulsar where the matter velocity
is negligible. However, as known, in case of a pulsar, neutrinos are
typically emitted from a neutrinosphere, which has a significant radius
especially for a young and hot star. The surface of the neutrinosphere
already has a nonzero velocity. Therefore the description of neutrinos
in a rotating matter proposed in ref.~\cite{BalPopStu09} is not
applicable in this case. In our approach, the quantum numbers $(N,s,p_{z})$,
characterizing a neutrino state, are measured by a corotating observer.
Thus our description can be used even when a neutrino is created in
a rotating matter, with all the noninertial effects being accounted
for exactly.

Nevertheless one can check that the energy spectrum in eq.~(\ref{eq:Energylev})
coincides with that found in ref.~\cite{BalPopStu09} for $N\omega\ll p_{z}$.
It happens when a pulsar rotates slowly or neutrinos are emitted close
the rotation axis, with initial momenta directed along it. In the
case we get from eq.~(\ref{eq:Energylev}) that $E_{A}-2N\omega\approx g^{0}+\sqrt{p_{z}^{2}+4N\omega g^{0}}$,
where we keep the term linear in $N\omega$ inside the square root.
One can see that, besides the additive term $2N\omega$, which is
due to the noninertial effects, the obtained expression coincides
with that derived in ref.~\cite{BalPopStu09}.

In section~\ref{sec:OSC} we have generalized our results to include
various neutrino generations as well as mixing between them. We have
derived the effective Schr\"{o}dinger equation which governs neutrino
flavor oscillations. We have obtained the correction to the resonance
condition in background matter owing to the matter rotation. Studying
neutrino oscillations in a millisecond pulsar, we have obtained that
the effective number density changes by $0.1\thinspace\%$.

Despite the obtained correction is small, we may suggest that our
results can have some implication to the explanation of great peculiar
velocities of pulsars. It was suggested in ref.~\cite{KusSeg96}
that an asymmetry in neutrino oscillations in a magnetized pulsar
can explain a great peculiar velocity of the compact star. An evidence
for the alignment of the rotation and velocity vectors of pulsars
was reported in ref.~\cite{Joh05}. Therefore we may suggest that
neutrino flavor oscillations in a rapidly rotating pulsar can contribute
to its peculiar velocity. It should be noted that neutrino spin-flavor
oscillations, including noninertial effects, in a rapidly rotating
magnetized star were studied in ref.~\cite{Lam05} in the context
of the explanation of high peculiar velocities of pulsars.


\acknowledgments{
I am thankful to S.~P.~Gavrilov for helpful comments, to FAPESP (Brazil) for a grant, to Tomsk State University Competitiveness Improvement Program for a partial support, and to Y.~S.~Kivshar
for the hospitality at ANU where a part of the work was made.}


\begin{thebibliography}{50}

\bibitem{Bel14}
  G.~Bellini, L.~Ludhova, G.~Ranucci and F.~L.~Villante,
  \textit{Neutrino oscillations},
  \textit{Adv. High Energy Phys.}
  \textbf{2014} (2014) 191960
  [arXiv:1310.7858]. 

\bibitem{BleSmi13}
  M.~Blennow and A.~Yu.~Smirnov,
  \textit{Neutrino propagation in matter},
  \textit{Adv. High Energy Phys.}
  \textbf{2013} (2013) 972485
  [arXiv:1306.2903]. 

\bibitem{BroGiuStu12}
  C.~Broggini, C.~Giunti and A.~Studenikin,
  \textit{Electromagnetic properties of neutrinos},
  \textit{Adv. High Energy Phys.}
  \textbf{2012} (2012) 459526
  [arXiv:1207.3980]. 

\bibitem{AhlBur96}
  D.~V.~Ahluwalia and C.~Burgard,
  \textit{Gravitationally induced neutrino-oscillation phases},
  \textit{Gen. Rel. Grav.}
  \textbf{28} (1996) 1161
  [gr-qc/9603008].

\bibitem{Lam01}
  G.~Lambiase,
  \textit{Neutrino oscillations in non-inertial frames and the violation of the equivalence principle. Neutrino mixing induced by the equivalence principle violation},
  \textit{Eur. Phys. J.} 
  \textbf{C 19}  (2001) 553.

\bibitem{DvoDib10}
  M.~Dvornikov and C.~O.~Dib,
  \textit{Spin-down of neutron stars by neutrino emission},
  \textit{Phys. Rev.} 
  \textbf{D 82} (2010) 043006
  [arXiv:0907.1445]. 

\bibitem{BasCho13}
  B.~Basu and D.~Chowdhury,
  \textit{Inertial effect on spin orbit coupling and spin transport},
  \textit{Ann. Phys. (N.Y.)} 
  \textbf{335} (2013) 47
  [arXiv:1302.1063].

\bibitem{GiuKim07}
  C.~Giunti and C.~W.~Kim,
  \textit{Fundamentals of Neutrino Physics and Astrophysics},
  Oxford University Press, Oxford (2007),
  pgs.~137--179.

\bibitem{DvoStu02}
  M.~Dvornikov and A.~Studenikin,
  \textit{Neutrino spin evolution in presence of general external fields},
  \textit{JHEP}
  09 (2002) 016
  [hep-ph/0202113].

\bibitem{QiaWan14}
  X.~Qian and W.~Wang,
  \textit{Reactor neutrino experiments: $\theta_{13}$ and beyond},
  \textit{Mod. Phys. Lett.} 
  \textbf{A 29} (2014) 1430016
  [arXiv:1405.7217]. 

\bibitem{Pet13}
  S.~T.~Petcov,
  \textit{The nature of massive neutrinos},
  \textit{Adv. High Energy Phys.}
  \textbf{2013} (2013) 852987
  [arXiv:1303.5819]. 

\bibitem{GriMamMos80}
  A.~A.~Grib, S.~G.~Mamaev and V.~M.~Mostepanenko,
  \textit{Quantum Effects in Intense External Fields:
  Methods and Results not Related to the Perturbation Theory},
  Atomizdat, Moscow (1980),
  pgs.~13--15.

\bibitem{PirRoyWud96}
  D.~P\'{\i}riz, M.~Roy and J.~Wudka,
  \textit{Neutrino oscillations in strong gravitational fields},
  Phys. Rev. 
  \textbf{D 54} (1996) 1587
  [hep-ph/9604403].

\bibitem{LanLif94}
  L.~D.~Landau and E.~M.~Lifshitz,
  \textit{The Classical Theory of Fields},
  Butterworth Heinemann, Amsterdam (1994), 4th edn.,
  pgs.~329--330.

\bibitem{Bak13}
  K.~Bakke,
  \textit{Rotating effects on the Dirac oscillator in the cosmic string spacetime},
  \textit{Gen. Relativ. Grav.}
  \textbf{45} (2013) 1845
  [arXiv:1307.2847]. 

\bibitem{SchWieGre83}
  P.~Schluter, K.-H.~Wietschorke and W.~Greiner,
  \textit{The Dirac equation in orthogonal coordinate systems:
  I. The local representation},
  \textit{J. Phys.} 
  \textbf{A 16} (1983) 1999.

\bibitem{ItzZub80}
  C.~Itzykson and J.-B.~Zuber,
  \textit{Quantum Field Theory},
  McGraw-Hill, New York (1980),
  pgs.~691--696.

\bibitem{HehNi90}
  F.~W.~Hehl and W.-T.~Ni,
  \textit{Inertial effects of a Dirac particle},
  \textit{Phys. Rev.} 
  \textbf{D 42} (1990) 2045.


\bibitem{Dvo11}
  M.~Dvornikov,
  \textit{Field theory description of neutrino oscillations},
  in \textit{Neutrinos: Properties, Sources and Detection},
  J.~P.~Greene eds.,
  Nova Science Publishers, New York (2011),
  pgs.~23--90
  [arXiv:1011.4300]. 

\bibitem{Dvo11b}
  M.~Dvornikov,
  \textit{Neutrino flavor oscillations in rotating matter},
  \textit{Azerbaij. Astron. J.}
  \textbf{6} (2011) 5
  [arXiv:1001.2516]. 

\bibitem{BalPopStu09}
  I.~Balantsev, Yu.~Popov and A.~Studenikin,
  \textit{Neutrino magnetic moment and neutrino energy quantization in rotating media},
  \textit{Nuovo Cim.} 
  \textbf{C 32} (2009) 53
  [arXiv:0906.2391]. 

\bibitem{LobStu01}
  A.~E.~Lobanov and A.~I.~Studenikin,
  \textit{Neutrino oscillations in moving and polarized matter under the influence of electromagnetic fields},
  \textit{Phys. Lett.} 
  \textbf{B 515} (2001) 94
  [hep-ph/0106101].

\bibitem{KusSeg96}
  A.~Kusenko and G.~Segr\`{e},
  \textit{Velocities of pulsars and neutrino oscillations},
  \textit{Phys. Rev. Lett.}
  \textbf{77} (1996) 4872
  [hep-ph/9606428].

\bibitem{Joh05}
  S.~Johnston, G.~Hobbs, S.~Vigeland, M.~Kramer, J.~M.~Weisberg and A.~G.~Lyne,
  \textit{Evidence for alignment of the rotation and velocity vectors in pulsars},
  \textit{Mon. Not. Roy. Astron. Soc.}
  \textbf{364} (2005) 1397
  [astro-ph/0510260].

\bibitem{Lam05}
  G.~Lambiase,
  \textit{Pulsar kicks induced by spin flavor oscillations of neutrinos in gravitational fields},
  \textit{Mon. Not. Roy. Astron. Soc.}
  \textbf{362} (2005) 867
  [astro-ph/0411242].

\end{thebibliography}
\end{document}